\documentclass[twocolumn,showpacs,preprintnumbers,amsmath,amssymb]{revtex4}
\usepackage{graphicx}
\usepackage{dcolumn}
\usepackage{bm}
\usepackage{epsfig}
\usepackage{subfigure}

\begin{document}

\title{Dynamical Mean Field Theory of Temperature and Field Dependent Band
Shifts in Magnetically Coupled Semimetals: Application to $EuB_6$}

\author{Chungwei~Lin and Andrew J.~Millis}

\affiliation{ Department of Physics, Columbia University \\
538W 120th St NY, NY 10027}

\begin{abstract}
A model for semimetals such as $EuB_6$, in which band overlaps are
controlled by magnetic order, is presented and is solved in the
dynamical mean field approximation. First order phase boundaries
are computed by evaluating free energies of different states. The
phase diagram is determined. A specific and physically reasonable
choice of parameters is found to approximately reproduce the
available data on $EuB_6$. For this material, predictions are made
for the location of a metamagnetic transition and its associated
endpoint, and a change in the order of the magnetic transition.
\end{abstract}


\maketitle

\section{Introduction}

$EuB_6$ is a magnetic semiconductor in which magnetic order
apparently controls the bandgap. The ferromagnetic transition at
$T_c \sim 12K$ is signaled by a sharp increase in plasma
frequency\cite{Degiorgy} and a sharp drop in
resistivity\cite{Aronson97} \cite{Aronson98} \cite{Ott00}. In this
paper we formulate a model which captures the physics of $EuB_6$,
and solve it in the dynamical mean field approximation. The
important technical step is the accurate determination of first
order phase boundaries via the construction of the free energies
of competing phases. We present a general phase diagram, which
includes a change in the order of the transition from first to
second. Remarkably, the change occurs not by the vanishing of a
fourth order coefficient but simply by an exchange of stabilities.
We show that for a choice of parameters consistent with what is
calculated for $EuB_6$ \cite{Pickett}, the various experimental
data are semiquantitatively reproduced. For these parameters, we
predict that $EuB_6$ should exhibit a metamagnetic phase
transition, and we estimate the location of the transition line
and its endpoints. The change in order of the phase transition
might be accessible in pressure experiments. The analysis reported
here may be thought of as a more precise solution of a model
proposed by \cite{Pickett}. Our solution agrees with theirs in
essential aspects but provides a sharper picture of physics and
includes some new features.

We also note that the band structure of the material is presently
the subject of controversy with photoemission experiments
\cite{Allen02} indicating a large gap (implying that the material
is intrinsically insulating so the metallic behavior is due to
defects which dope the system) while some band calculations
\cite{Massidda} \cite{Pickett} and quantum oscillation
measurements\cite{Aronson99} indicating a small negative gap
causing the semimatal behavior. We comment below on the
implications of our work for these controversies.


\section{Model Hamiltonian}
The electronic structure of $EuB_6$ has been calculated
\cite{Massidda} \cite{Pickett}. Experiment, quantum chemical
intuition and band calculations all agree that the $Eu$ $f$-shell
is half filled and, as the scale relevant for electronic behavior,
electronically inert. Crystal field effects are negligible and to
a good approximation, each $Eu$ may be regarded as carrying a
$S=7/2$ ``core spin''. Band theory calculations \cite{Pickett}
reveal two near Fermi surface bands: a nearly empty band with a
minimum at X-point, derived mainly from $Eu-d$, $B-p$ orbitals,
and a nearly full band with a maximum at the X-point, derived
mainly from $Eu-f$, $B-p$ ones. The hybridization between these
bands is negligible because they arise from different symmetry
orbitals.

\begin{figure}[ttt]
\hspace*{-0.5cm}
\includegraphics[width=3in]{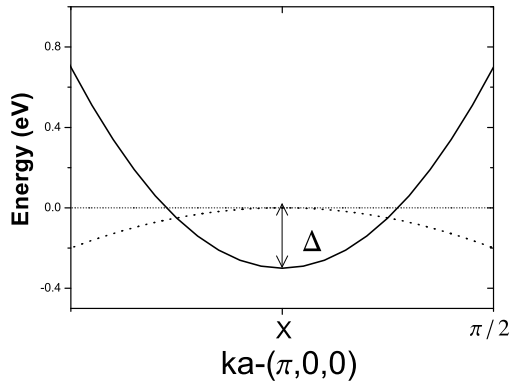}
   \caption{Expanded view of ferromagnetic phase band structure of $EuB_6$ with
    momentum $k$ measured from
    $X$ [$\pi,0,0$], in units of inverse lattice constant $a=4.2 A$.
    Curved (flat) dashed line represents valence (defect) band and heavy line conduction band.}
   \label{fig:band}
\end{figure}

Because we shall be interested in low energies and weak couplings, we expand the
bands near the X-point so the minimal model describing $EuB_6$ becomes:
\begin{eqnarray}
H&=& \sum_{\sigma,\vec{p}} \left(\frac{p^2}{2\,m_1}+\triangle \right) c^+_{1,\sigma, \vec{p}}
c_{1,\sigma,\vec{p}}-\frac{p^2}{2\,m_2} c^+_{2,\sigma, \vec{p}} c_{2,\sigma,\vec{p}}
\nonumber \\&&
+ \sum_{i,a,b, \alpha,\beta} J^{a,b} \vec{S}_i \cdot
c^+_{a,i,\alpha}\, \vec{\sigma}_{\alpha \beta}\, c_{b,i,\beta} \nonumber \\ &&
-\mu \sum_i (n_{i,1}+n_{i,2})
\label{eqn:H}
\end{eqnarray}
Here $\vec{S}$ represents the $S=7/2$ $Eu$ core spin, $c_1$ and
$c_2$ represent upper (conduction) and lower (valence) bands and
$\triangle$ controls the band overlap. We take $|\vec{S}|=1$,
absorbing the actual magnitude into the coupling $J$. Because
$\vec{S}$ originates from a filled (spin polarized) $Eu$ shell,
and the latest band structure calculation \cite{Pickett} shows
that two bands arises from different $Eu$ orbitals ($f$ and $d$),
we may therefore take $J$ to be diagonal in orbital indices. The
high spin and low carrier density means the Kondo effect is
irrelevant, so the sign of $J$ is arbitrary. Kunes and Pickett's
calculation \cite{Pickett} implies $J^{11}<0$ (Kondo coupling) and
$J^{22}>0$ (anti-Kondo coupling) and indicates that the splitting
of the conduction and valence bands are roughly the same, we
choose $J=J^{22}=-J^{11}$. Our results turn out to depend mainly
on $|J^{22}|+|J^{11}|$. The chemical potential $\mu$ is determined
by the neutrality of the system, i.e. number of particles equals
number of holes. Fig(\ref{fig:band}) shows the simplified band
structure implied by eqn(\ref{eqn:H}).

The model described in eqn(\ref{eqn:H}) can be easily modified to
match the bandstructure implied by the photoemission
experiment\cite{Allen02} as follows: we interpret the lower band
($c_2$) as arising from defect states which in this picture must
exist so that carriers are donated to the conduction band, and
thus take $m_2$ to be infinite, and assure there is no coupling
between $c_2$ band and the core spin $\vec{S}$ ($J^{22}=0$). At
these parameters ($m_2 \rightarrow \infty$, $J^{22}=0$), the $c_2$
band serves as a particle reservoir and the chemical potential is
fixed at 0. This choice in fact does not change the structure of
the phase diagram or the qualitative features of our results.

The $T=0$ phase diagram is straightforward. The ground state is ferromagnetic. If
$\triangle+|J^{11}|+|J^{22}|<0$, the majority spin of band 1 crosses the minority
spin of band 2, which results in metallic behavior. If not, the ground state is
insulating. At $T>T_c$, the random spin orientations mean that the band shift are
much less (especially at small $J$). Roughly, $\triangle<0$ implies metallic behavior
while $\triangle>0$ implies insulating state.


\section{Basic Results}

\subsection{Dynamical Mean Field and Free Energy}
To study eqn(\ref{eqn:H}), we use the single site dynamical mean
field method \cite{Georges}. This amounts to assuming that the electron self energy is
momentum independent: $\Sigma_{1,\sigma}(\vec{p},\omega)\rightarrow
\Sigma_{1,\sigma}(\omega)$. The self energy is determined from the solution of
an auxiliary impurity model with free energy $\Omega_{imp}=-T \log Z_{imp}$, and
\begin{eqnarray}
Z_{imp}=\int \,D[c^+ c]\,D[\vec{S}_c]\, \exp\,S_{eff} \nonumber
\end{eqnarray}
with \cite{coupling}
\begin{eqnarray}
S_{eff}&=&\frac{1}{\beta^2}\int_0^{\beta} d\tau d\tau' \bar{c_{\alpha}}(\tau) a_{\alpha \beta}
 (\tau-\tau') c_{\beta}(\tau')  \nonumber \\ &&+
 \frac{1}{\beta} \int d\tau  J \vec{S} \cdot \left[ \vec{\sigma}^{\alpha \beta}
 \bar{c}_{\alpha}(\tau) c_{\beta}(\tau) +h\hat{z} \right]
\label{eqn:Zeff}
\end{eqnarray}
The free energy of the lattice model is \cite{Georges}
\begin{eqnarray}
&&\frac{\Omega}{N}=\Omega_{imp}-T \sum_{n, \sigma}\log
G_{\sigma}(i \omega_n) \nonumber \\  &&-T\sum_{n,\sigma}  \int d
\epsilon D(\epsilon) \times \log [i
\omega_n+\mu-\Sigma_{\sigma}(i\omega_n)-\epsilon]
\label{eqn:LatFreeEnergy}
\end{eqnarray}
and the auxiliary function $a$ is fixed by the requirement
$\frac{\partial \Omega} {\partial a}=0$. It is important to
perform the exact quantum trace over the spin degrees of freedom
rather than employing the classical spin approximation often made
\cite{millis} in order to obtain physically reasonable estimates
for the entropy.

The free energy so determined is a functional of the applied field
$h$ and the system magnetization is given by $m=-\frac{\partial
\Omega}{\partial h}$. It is sometimes useful to perform a Legendre
transformation to obtain $\Omega(m,T)$. We have not found an
efficient method for performing this transformation: to obtain
$\Omega(m,T)$, we solve the dynamical mean field equations
numerically for a range of $h$, and then construct
$\Omega(m,T)=\Omega(h,T)+h\,m$ explicitly. We note however that in
some parameter ranges, the DMFT equations have two stable
(symmetry un-related) solutions at $h=0$, corresponding to states
with $m=0$ and $m \neq 0$. The two states so determined are
extrema of $\Omega(m,T)$ and their free energies are the extremal
values of $\Omega(m,T)$, so first order transition points may be
located without constructing $\Omega(m,T)$ explicitly.


\subsection{Approximations and DMFT Solutions}
We have solved the dynamical mean field equations corresponding to
extremizing eqn(\ref{eqn:LatFreeEnergy}) with respect to
$\Sigma(\omega_n)$. For simplicity we adopted a semicircular
density of states $D(\epsilon)=\frac{\sqrt{4t^2-\epsilon^2}}{2 \pi
t^2}$ with $t=\frac{(\sqrt{2} \pi)^{\frac{1}{3}}}{m a^2} $ chosen
to match the band theory band-edge density of states. We also
assumed $J/t<<1$ (because this is the limit relevent for the
materials of interest) and $T/J<<1$ (we will see below that the
calculated transition temperature is much less than $J$). These
approximations simplify calculations considerably, in particular,
in this limit we may retain only the $z$ component of the core
spin, simplifying the quantum trace.

To simplify the expression, we define
\begin{eqnarray}
\triangle= 2 y J
\end{eqnarray}
where $y$ is a dimensionless parameter measuring the bandgap. We
also define the following $f$ functions:
\begin{eqnarray}
f^{15}(x)&=&\Theta(x)x^{1.5}\nonumber \\
f^{25}(x)&=&\Theta(x)x^{2.5}\nonumber
\end{eqnarray}
where $\Theta(x)$ is the step function.

In the small $J$ limit,
\begin{eqnarray}
\Sigma_{\uparrow,(\downarrow)}&=&\mp mJ
\end{eqnarray}
the free energy $\triangle \Omega$ and core spin magnetization $m$ are given by
\begin{eqnarray}
&&\triangle \Omega(m) = -\frac{4\,J}{15\pi} \times \nonumber \\
&& \left[ (\frac{J}{t_1})^{3/2} \left(f^{25}(\mu-(y-m))+
f^{25}(\mu-(m+y)) \right) \right. \nonumber \\ && \left. +
(\frac{J}{t_2})^{3/2} \left( f^{25}(m-y-\mu)+f^{25}(-(m+y)-\mu) \right) \right] \nonumber \\
&& +T\, m \alpha
-T \log \left[\frac{\sinh[(1+ \frac{1}{2S})\alpha]}{\sinh[\frac{\alpha}{2S}]} \right]
\label{eqn:Omega(m)}
\end{eqnarray}
\begin{eqnarray}
m= \frac{8}{7}\sinh(\frac{8\alpha}{7})-\frac{1}{7}\sinh(\frac{\alpha}{7})
=B_{7/2}\left[\alpha(m,\beta)\right]
\label{eqn:m=f(m)}
\end{eqnarray}
where
\begin{eqnarray}
&& \alpha = \frac{2}{3 \pi} \beta J  \times \nonumber \\
&& \left[ \left( \frac{J}{t_1} \right)^{1.5}
\left(f^{15}(\mu-(y-m))- f^{15}(\mu-(m+y)) \right) \right. +
\nonumber
\\ && \left. \left( \frac{J}{t_2} \right)^{1.5} \left(
f^{15}(m-y-\mu)-f^{15}(-(m+y)-\mu) \right) \right]
\end{eqnarray}

The chemical potential $\mu$ is determined by charge neutrality, i.e.
number of electron equals number of holes.
\begin{eqnarray}
&&\frac{1}{t_1^{1.5}} \left(f^{15}(\mu-(m-y))+ f^{15}(\mu-(m+y))
\right) \nonumber \\ &=& \frac{1}{t_2^{1.5}} \left(
f^{15}(m-y-\mu)+f^{15}(-(m+y)-\mu) \right) \label{eqn:mu}
\end{eqnarray}
Note that $\mu$ is a function of $m$ and $y$, not an independent
variable.

Based on the above expressions, the entropy $S$ is
\begin{eqnarray}
S=-\frac{\partial \triangle \Omega}{\partial T}
=\log \left[\frac{\sinh( \frac{8\alpha}{7})}{\sinh(\frac{\alpha}{7})} \right]
-m \alpha
\end{eqnarray}
the specific heat $C_V$ is
\begin{eqnarray}
C_V=\frac{\partial\, \triangle E}{\partial T}=\frac{\partial
(\triangle \Omega+TS)} {\partial T} = -T \alpha \frac{\partial
m}{\partial T}
\end{eqnarray}
$\frac{\partial m}{\partial T}$ can be computed from eqn(\ref{eqn:m=f(m)}).

Finally, the density of conducting band electrons $\triangle n$ is
\begin{eqnarray}
&& \triangle n = \frac{4}{3 \pi} (\frac{J}{t_1})^{3/2}
(f^{15}(\mu(m,y)-(y-m))+\nonumber \\
&&f^{15}(\mu(m,y)-(m+y))-2\,f^{15}(\mu(0,y)-y) )
\end{eqnarray}
The plasma frequency $\omega_{p}$ is defined as $\omega_{p}^2=4 \pi e^2\frac{n}{m^*}$ and
calculated by
\begin{eqnarray}
\omega_{p}^2(T_c^-)-\omega_{p}^2(T_c^+)=4 \pi e^2\frac{\triangle n}{2}
\left( \frac{1}{m_1^*}+\frac{1}{m_2^*} \right)
\end{eqnarray}

We emphasize that all expressions above should be evaluated at the $m$ where
eqn(\ref{eqn:m=f(m)}) is satisfied.

\begin{figure}[ttt]
\hspace*{-0.5cm}
\includegraphics[width=3in]{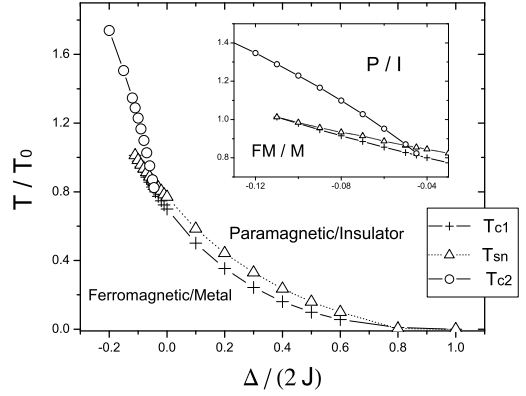}
   \caption{The phase diagram. In the range $-0.1J<\triangle<-0.55J$, the system undergoes both
    1st and 2nd order phase transitions. The temperature is in unit
    $\frac{2}{3 \pi} t_1 (\frac{J}{t_1})^{2.5}$. $T_0=17.43\,K$ if the $J=0.16 eV$ and $t_1=1.16eV$
    implied in Kunes/Pickett's bandstructure calculation.}
   \label{fig:PhaseDia}
\end{figure}

The calculated phase diagram is shown in Fig(\ref{fig:PhaseDia}). We see, in accord with
the simple considerations of the previous section that for $\triangle / J>1$, the material
is always insulating and $T_c$ is negligible. For $\triangle$ sufficiently negative, the
transition becomes second order.

\begin{figure}[htbp]
   \centering
   \subfigure[$\triangle=0$]{\epsfig{file=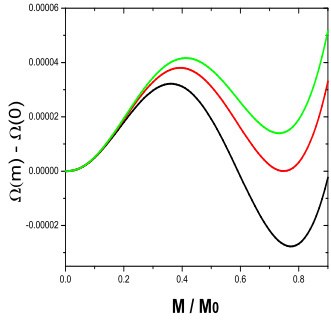, height=2in, width=2in}}
   \subfigure[$\triangle=-0.14J$]{\epsfig{file=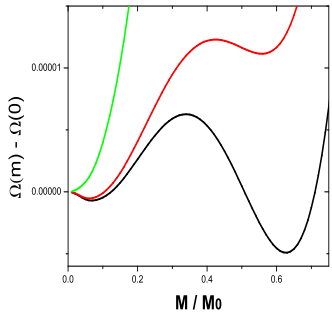, height=2in, width=2in}}
   \subfigure[$\triangle=-0.3J$]{\epsfig{file=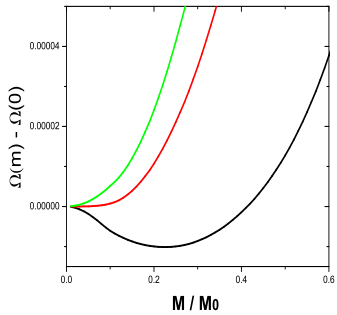, height=2in, width=2in}}
   \caption{Free Energy as a function of $m$ at different temperature. (a) $\triangle=0$ where
    only first order phase transition occurs. From top to bottom:
    $T>T_{c1}$, $T=T_{c1}$, and $T<T_{c1}$. (b) $\triangle=-0.14J$ where both first and
    second phase transition occur. From top to bottom:  $T=T_{c2}$,
    $T_{c2} \, > \,T \, > \, T_{c1}$,
    and $T<T_{c1}$. (c) $\triangle=-0.3 J$ her only
    second order phase transition occurs.
    From top to bottom: $T>T_{c2}$, $T=T_{c2}$, and $T<T_{c2}$.}
   \label{fig:FE(m,T)}
\end{figure}

\begin{figure}[ttt]
\hspace*{-0.5cm}
\includegraphics[width=3in]{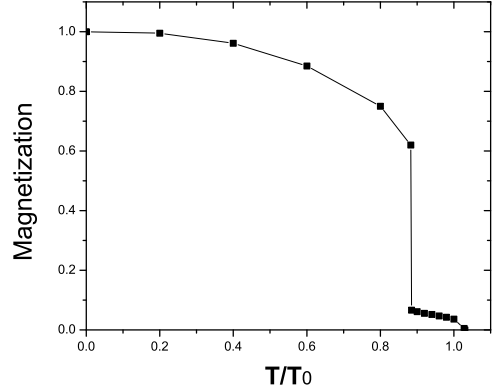}
   \caption{The magnetization $m$ in units of the $T=0$ saturation magnetization $m_0$
    as a function of temperature at $\triangle=-0.14J$. $T_0 =17.43\,K$.
    The second order transition at $T \approx 1.028 \,T_0$, followed by the
    first order transition at $T \approx 0.89 \,T_0$. }
   \label{fig:MvsTNegY}
\end{figure}

Fig(\ref{fig:FE(m,T)}) shows the the free energy as a function of
$m$ at several temperatures and at different bandgaps. Panel (a)
shows $\Omega(m)$ for several $T$ at $\triangle/J=0$, in the first
order region of the phase diagram. Panel b shows the unusual
behavior in the vicinity of the multicritical point where the
transition becomes second order. One naively expects the minimum
at higher $m$ shifts downwards, eventually merging with $m=0$
minimum (in other word, that the sign of the $m^4$ term in the
Landau expansion of the free energy changes). Panel (b) shows that
the actual situation is more subtle: the higher $m$ minimum
continues to exist; however its energy is increased so the second
order phase transition happens, and then at a lower temperature
the higher minimum takes over. The temperature dependence of the
magnetization in this region is shown in Fig(\ref{fig:MvsTNegY}).
Finally, panel (c) shows the free energy in the region in which
there is no first order transition at all. We note that pressure,
by increasing hybridization, is expected to broaden bands, thereby
makes $\triangle$ more negative. The predicted  change in order of
transition may therefore be accessible in pressure experiments.

\begin{figure}[ttt]
\hspace*{-0.5cm}
\includegraphics[width=3in]{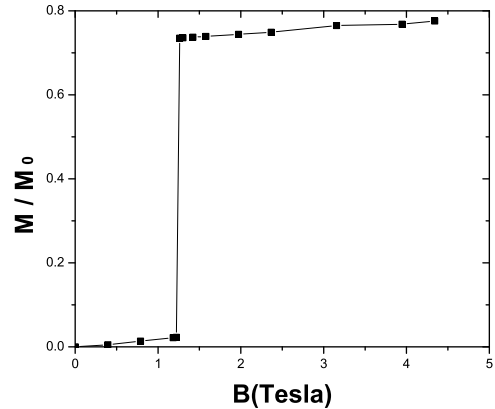}
   \caption{The magnetization $m$ in units of the $T=0$ saturation magnetization $m_0$
    in the paramagnetic phase at $T=12.55\, K$ ($T_c=12.2\,K$),
    as a function of magnetic field $B$, calculated from minimization of free energy
    for parameters $t_1=1.16\, eV$, $t_2=0.68\, eV$, $J=0.16 \,eV$, appropriate to $EuB_6$. }
   \label{fig:M(h)}
\end{figure}

The presence, in the wide regions of the phase diagram, of a metastable high $m$
free energy minimum, suggests that the system may undergo a metamagnetic phase transition.
As shown in Fig(\ref{fig:M(h)}), this is indeed the case.

\section{Application to $EuB_6$}

In this section, the model is applied to $EuB_6$. The literature
presents two conflicting interpretations of this compound: the
semimetal interpretation favored by quantum
oscillations\cite{Aronson99} and bandstructure
calculations\cite{Pickett}, and the large band gap interpretation,
implied by photoemmsion\cite{Allen02}. These imply quite different
parameters; we consider them seperately, beginning with the
semimetal case.

There are four parameters in this model -- conduction and valence
bandwidths $t_1$ and $t_2$, coupling strength $J$, and the bandgap
$\triangle$. $t_1$ and $t_2$ are fixed by the effective masses of
both bands given in \cite{Pickett}, and we found $t_1=1.16\,eV$,
$t_2=0.68\, eV$. Two sets of $J$ and $\triangle$ are chosen --
$J=0.14 \, eV $, $\triangle=-0.2 J$ and $J=0.16 \, eV
$,$-0.04\,J<\triangle<0.04\,J$. The former choice are the best-fit
parameters for fixed critical temperature and plasma frequency
jump. However, these parameters produce a specific heat in
disagreement with experiment and are in the region where the
system has both 1st and 2nd order transitions whereas experiment
apparently yields only one transition \cite{Aronson98} We cannot
fit $T_c$ and $\triangle \omega_p^2$ simultaneously in the 1st
order transition region. $J=0.16\,eV$,
$-0.04\,J<\triangle<0.04\,J$ are chosen the match the critical
temperature, but the calculated $\triangle \omega_p^2$ and
$\triangle C_V$ are roughly 2.5 and 2 times larger than the
experimental data.

With the two sets of parameters, we compute critical temperature
$T_c$, jump in plasma frequency $\triangle \omega_p^2$, jump in
specific heat $\triangle C_V$, core spin magnetization $m$, and
latent heat $\triangle Q$. The results are summarized and compared
with experiment in the following table.

\begin{center}
\begin{tabular}{|c|c|c|c|} \hline
  Quantity  & Expt & (0.14,-0.2$J$) & (0.16, $\pm 0.04 J$)     \\ \hline
$T_c$ ($K$) & 12$\sim$14 & 12.2  &$12.2 \pm 1$ \\ \hline
$\triangle \omega_p^2$ ($10^{7}cm^{-2}$) & 1.625 &  1.7 &
$4.3\pm0.2$\\ \hline \begin{tabular}{c} $\triangle C_V$ \\ ($J/K$
per mole) \end{tabular} & 12&
\begin{tabular}{c}69.63 \\ (10.47$\rightarrow$80.1)
\end{tabular} & $21.2\pm 2.4$ \\ \hline $m$
/ $m_0$ & * & \begin{tabular}{c}0.3 \\(0.28$\rightarrow$0.58)
\end{tabular} & $0.75\pm 0.5$  \\
\hline $\triangle Q$ ($J/Mole$) & * & 30.92 & $90.5 \pm 2$ \\
\hline
\end{tabular}
\end{center}
where the parentheses in the table represents $(J (eV), \triangle
)$. For the parameters $J=0.14 \, eV $ and $\triangle=-0.2 J$,
$T_{c2}=16.1\,K$.

\begin{figure}[ttt]
\hspace*{-0.5cm}
\includegraphics[width=3in]{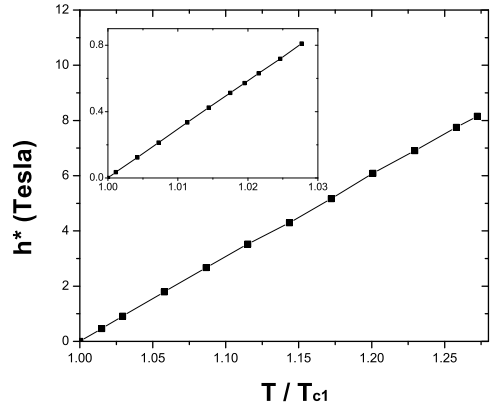}
   \caption{$h^*(T)$, see the text for the definition of $h^*$. The matemagnetic phase
    transition ends at $T=1.28 \, T_{c1}$ for $J=0.16 eV$, $\triangle=0$.
    The inset is for parameters $J=0.2 eV$, $\triangle=-0.2J$ where the
    metamagnetic transition ends at $T=1.01 T_{c1}$. }
   \label{fig:h^*(T)}
\end{figure}

The existence of a metastable $m\neq 0$ state implies the
existence of a metamagnetic phase transition. For example,
Fig(\ref{fig:M(h)}) shows $M(B)$ calculated at a $T$ slightly
greater than $T_c$. We take $T= 12.55 K$ ($T_c=12.2 K$). We see
that the jump occurs at $B=1.12$ tesla, the magnetic field where
the magnetization jump occurs as $h^*$, Fig(\ref{fig:h^*(T)})
plots $h^*$ as a function of temperature.

\begin{figure}[ttt]
\hspace*{-0.5cm}
\includegraphics[width=3in]{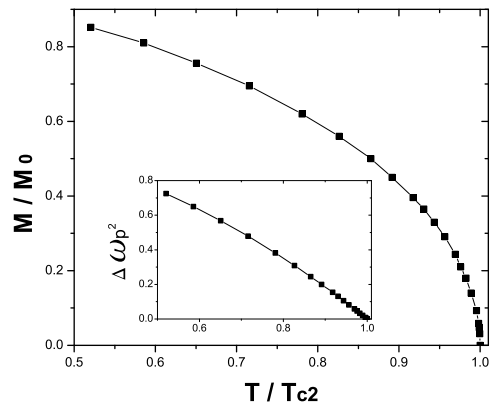}
   \caption{$M(T)$ at photoemission parameters. $T_{c2}=12.5K$.
    The inset shows the square of plasma frequency ($\triangle \omega_p^2$)
    (in the unit $10^7 cm^{-2}$) as a function of temperature.}
   \label{fig:1_orbital}
\end{figure}

Photoemission experiments\cite{Allen02} at $T>T_c$ detects a wide
range band gap ($\triangle \sim 1eV$), and a fermi level about
$0,2 eV$ above the conduction band minimum. The carriers in this
band must arise from defects which dope the system; we model the
defects as a flat band ($m_2\rightarrow\infty$, $t_2=0$). The
measured conduction band dispersion implies $t_1=1.54eV$; the
fermi level position yields $\triangle=-0.2 eV$; $J^{22}=0$, and
by fitting $T_c$ we obtain $J^{11}=0.14eV$ (quite close to the
band theory value). These parameters place the system well into
the second order region of the phase diagram in apparent
contradiction to experiments. In this case, the jump in specific
heat is $19.24$ (J/Mole) while the plasma frequency and
magnetization are all continuous as a function of temperature, as
shown in fig(\ref{fig:1_orbital}). \

\section{Conclusion}
We proposed a model to explain the combined metal/insulator,
ferro/para magnetic phase transition of $EuB_6$. The splitting of
the $Eu$ derived conduction band due to the coupling to core
electrons at each $Eu$ site is found to be the origin of the phase
transition. From the technical point of view, the new feature of
our results is the detailed examination of the dynamical mean
field free energy. The free energy calculation implies the
transition is first order. By choosing the physically reasonable
parameters, the available experimental data ($T_c$, jump in plasma
frequency, and jump in specific heat) are reasonably well
reproduced if the picture is favored by band theory and
transported is adopted. The calculated jump in specific heat is
around two times greater than the experiments. We predict a
metamagnetic phase transition at fields up to $8$ tesla, if the
temperature is within a few degrees of $T_{c1}$. We predict that
the modest pressures will change the order of phase transition
from first to second. The parameters found by photoemission lead
to a second order transition, a disagreement with other
experiments.

This work is supported by University of Maryland-Rutgers MRSEC.




\begin{thebibliography}{1}
\bibitem{coupling}
We included a magnetic field coupling only to the core spin because
we only want to express the free energy as a function of the core spin magnetization,
not the total magnetization. In fact, if one wants to express the free energy
as function of total magnetization, our answer should be a good approximation
since the $g$ factor is 1/2 for free electron while 7/2 for core spin.
\end{thebibliography}
\end{document}